# Excitation and detection of acoustic phonons in nanoscale systems

*Ryan C. Ng*[*,1], *Alexandros El Sachat*[*,2], *Francisco Cespedes*[1,3], *Martin Poblet*[1], *Guilhem Madiot*[1], *Juliana Jaramillo-Fernandez*[1], *Peng Xiao*[1,3], *Omar Florez*[1,3], *Marianna, Sledzinska*[1], *Clivia Sotomayor-Torres*[1,4], *Emigdio Chavez-Angel*[*,1]*

[1] *Catalan Institute of Nanoscience and Nanotechnology (ICN2), CSIC and BIST, Campus UAB, Bellaterra, 08193 Barcelona, Spain*

[2] *National Center for Scientific Research "Demokritos", 15310 Athens, Greece*

[3] *Departamento de Física, Universidad Autónoma de Barcelona, Bellaterra, 08193 Barcelona, Spain*

[4] *ICREA, Passeig Lluis Companys 23, 08010 Barcelona, Spain*

\* *ryan.ng@icn2.cat, alexandros.elsachat@icn2.cat, emigdio.chavez@icn2.cat*

## Abstract

Phonons play a key role in the physical properties of materials, and have long been a topic of study in physics. While the effects of phonons had historically been considered to be a hindrance, modern research has shown that phonons can be exploited due to their ability to couple to other excitations and consequently affect the thermal, dielectric, and electronic properties of solid state systems, greatly motivating the engineering of phononic structures. Advances in nanofabrication have allowed for structuring and phonon confinement even down to the nanoscale, drastically changing material properties. Despite developments in fabricating such nanoscale devices, the proper manipulation and characterization of phonons continues to be challenging. However, a fundamental understanding of these processes could enable the realization of key applications in diverse fields such as topological phononics, information technologies, sensing, and quantum electrodynamics, especially when integrated with existing electronic and photonic devices. Here, we highlight seven of the available methods for the excitation and detection of acoustic phonons and vibrations in solid materials, as well as advantages, disadvantages, and additional considerations related to their application. We then provide perspectives towards open challenges in nanophononics and how the additional understanding granted by these techniques could serve to enable the next generation of phononic technological applications.

## Introduction

Pioneering studies of confined acoustic waves were first performed by Lord Rayleigh[1] in 1885. He demonstrated the existence of surface acoustic waves (SAWs) propagating along the surface of an isotropic solid half-space. These waves are non-dispersive and elliptically polarized in the sagittal plane, with velocities that are slightly smaller than that of bulk shear waves, and with amplitudes that decay exponentially, confining them to roughly within a wavelength of the surface. Following his results, other scientists further developed this topic, particularly: Stonely,[2] who described waves that propagate along the interface between two half-spaces, i.e., a generalization of Rayleigh modes; Love,[3] who described horizontally polarized surface waves; and Sewaza,[4] who demonstrated the existence of surface waves in layered structures. The propagation of SAW modes in these types of systems has been widely studied by the seismology community, who were the first to study the confinement of acoustic waves, although their relevance is pervasive and extends throughout modern devices.[5] Years later, the acoustic properties of other finite systems and geometries were also studied. Pochhammer[6] and Chree[7] described the propagation of waves in infinite cylindrical rods. Lamb developed theoretical models to describe the natural vibration of spheres[8] and suspended thin films[9]. While these demonstrations were originally applied in "bulk-like" or macro-scale systems, the results are scale invariant and still hold down to the nano-[10] and molecular[11] scales.

The quantization of the vibrational energy of an atom or a group of atoms in matter is a quasiparticle known as a phonon. As phonons are quasiparticles, each phonon obeys Bose-Einstein statistics and has an energy $\hbar\omega$ and a pseudo-momentum $p = \hbar q$, where $\omega = 2\pi f$, $f$ is the phonon frequency, $\hbar$ is the reduced Planck constant ($h/2\pi$), and $q$ is the wavevector. Similar to other quasiparticles, the wavelength dependence of the phonon energy can be represented by a dispersion relation, or a relationship between the phonon frequency and its wavevector. The phonon group velocity, $v_g = d\omega/dq$, can be determined from the slope of a dispersion relation curve. For bulk material, the dispersion relation of acoustic phonons with short wavevector (i.e., near the $\Gamma$-point in the first Brillouin zone) is linear and the group velocity in the material is constant.[12] Upon decreasing the characteristic dimensions of the material, this linear dependence no longer holds, and the phonon energy is quantized into many discrete modes. This spatial confinement resulting from the reduction of bulk material down to "finite" nanometer sizes affects a wide array of physical properties such as the phonon density of states, group velocity, specific heat capacity, and electron-phonon and phonon-phonon interactions, among others.[13–18] Phonon statistics and the interaction of phonons with other particles sets a limit to properties such as the electrical and thermal conductivity, sound transmission, reflectivity of ionic crystals, the linewidth of semiconductor quantum dot emission, the maximum power that can be carried by optical fibres, and the inelastic scattering of light, x-Rays, and neutrons.[19,20] Thus, the engineering of new devices that are able to generate, control, and detect phonons becomes a key issue that is essential for the understanding and development of future technologies.

The elastic continuum model predicts that the acoustic/mechanical frequency scales inversely with the characteristic dimension of a structure. For example, the first natural mode of a solid sphere scales as $f = \frac{\theta}{\pi D}\sqrt{\frac{G}{\rho}}$ where $\theta$ is a root of a polynomial equation, $D$ is the diameter, $G$ is the shear modulus, and $\rho$ is the density.[21] As the size of the structure decreases, these modes increase in energy and can interact with particles or quasi-particles existing in the medium. Moreover, the decrease in size can also reduce the phonon mean free path due to diffusive scattering caused by the surface roughness of the boundaries. This has a direct impact on properties such as the thermal transport, where phonons are the main energy carriers in semiconductors and insulators. At the nanoscale, heat conduction is affected by: (*i*) the increase in boundary (interface) scattering and (*ii*) the effect of phonon confinement. At room temperature, the impact of phonon confinement on thermal transport is negligible. Rather, the modification of the thermal conductivity is mainly attributed to diffuse scattering of phonons at boundaries. Although this mechanism has been widely explored and exploited,[22–24] recent works propose using phonon coherence to control heat flow.[25–27] In this scenario, modification of the phonon dispersion relation by adding additional periodicity to the system by alternating thin layers of dissimilar materials (superlattices),[28] patterning holes in a suspended system,[29,30] or by self-assembling colloidal particles,[31,32] offers a simple strategy to manipulate phonon waves. These kinds of systems are known as phononic crystals (PnCs) or acoustic metamaterials.

The generation, manipulation, and detection of phonons are three central concepts that are addressed by the phononic community, and their appropriate incorporation and consideration are essential towards realizing the aforementioned applications. Compared to its photonic and electronic counterparts, phononics has received less attention, in part due to the difficulty associated with the experimental excitation and detection of acoustic phonons. Beyond characterization of phonons, even the development and fabrication of platforms that operate at higher energies (i.e., GHz) are challenging due to limitations in nanofabrication and the difficulties associated with pushing fabrication resolution to a few nanometer feature size. While phonons have traditionally been considered to be an inconvenience in most materials and devices, recent research has shown that phonons can be exploited for a wide variety of applications. Existing reviews have focused on thermal phonons[25,30,33–36] or have provided a holistic analysis of the interaction and transduction between photons, electrons, and phonons within electro- and opto-mechanical systems,[37,38] though a comparison of existing experimental methods available to probe and directly measure mechanical modes and acoustic phonons (i.e., low energy phonons, <THz) is still missing. Thus, we summarize the available experimental methods for the excitation and detection of acoustic phonons and their associated advantages, disadvantages, and additional considerations. We first review the advances in techniques that are capable of phonon detection (Raman spectroscopy and laser Doppler vibrometry) and then follow with techniques that are capable of both phonon excitation and detection (Brillouin light scattering, scanning probe microscopy, cavity optomechanics, pump-probe techniques, and interdigitated transducers). Within each section we also highlight interesting applications of each technique. Finally, we conclude by offering perspectives of the field.

## Detection Techniques

### Raman spectroscopy

Raman spectroscopy is a non-contact, optical characterization technique widely used for the elemental analysis of materials. This technique detects light scattered by atomic/molecular vibrations (i.e., phonons) of the system under study. As vibrations depend on the nature of atomic/molecular bonds and the specific atoms in the system, each vibrational mode provides a unique fingerprint allowing for chemical identification of different materials. This technique is most commonly used as a basic characterization tool to identify materials by comparing measured spectra with that of a database (see e.g., www.rruff.info or www.irug.org). However, the potential of Raman spectroscopy can go much further and additionally provide a plethora of information such as crystal orientation, amorphous domains, chemical stoichiometry, contamination traces, strain, ferroelectric and ferromagnetic domains, pseudo phases, and thermal properties, to name a few. In recent years, with the rapid development of low-dimensional material-based applications, Raman spectroscopy has been used to characterize the size and chemical functionalization of low-dimensional materials such as nanodots, nanowires[39], and two-dimensional (2D) layered materials[40].

These low-dimensional materials are interesting in nanophononics due to their ability to confine phonons as they approach length scales in which their finite size begins to have a strong effect on material properties. For example, in the case of nanoparticles, the confinement resulting from the edges of the nanoparticle leads to a break in translational symmetry in the crystal (i.e., absence of periodicity beyond the limits of the particle). This loss of symmetry leads to a shift of the optical Raman modes and an asymmetric broadening of the signal, usually observed as a "shoulder" at lower wavenumber. This is induced by the well-known phonon confinement effect. Historically, this effect has been described by a Gaussian confinement model (the Richter-Cambell-Fauchet RCF model),[41,42] an elastic[43,44] and dielectric[45] continuum model, microscopic lattice dynamic calculations for nanoparticles,[46] and Density-Functional Perturbation Theory (DFPT).[47] The finite size of low-dimensional materials is conducive to the activation of forbidden modes[48] or the detection of confined acoustic phonon modes.[49,50] In general, Raman spectroscopy can only detect zone-center optical phonons (i.e., $q = 0$). This selection rule is a consequence of the infinite periodicity of a crystal lattice.[51] However, in finite-sized systems in which the translational symmetry is broken, the zero-center optical phonon selection rule is relaxed. This causes the Raman spectrum in these finite systems to also have contributions from phonons that are further away from the Brillouin-zone center. This effect can be observed in systems smaller than ~20 lattice parameters.[51] Beyond this limit for larger systems, the contribution of this effect to the Raman spectra is negligible. Acoustic modes (i.e., the natural elastic vibrations of a system) are always present in materials, although they may be difficult to detect. As finite length scales are approached, acoustic modes can begin to be detected. For spherical nanoparticles, their finite size allows for the detection of several discrete Lamb's modes that are associated with spheroidal and torsional modes of the particle.[49,52–54] Early demonstrations of detection of phonon confinement using Raman scattering were performed by Colvard et al,[55] who measured the folding of longitudinal acoustic phonons in GaAs/AlAs superlattices. They also demonstrated that the phonon folding follows the well-known Rytov equation for infinitely long superlattices,[56] a model based on the elastic continuum theory. The first unambiguous demonstration of confinement of optical phonons was measured by Jusserand et al in 1984, detected using Raman spectroscopy in GaAs/GaAlAs superlattices.[43]

Confinement in superlattices can be considered to be one-dimensional (1D) since the confinement occurs along the layered axis. Other examples of 1D confinement occur in thin films and van der Waals layered materials. In thin films, confinement of both acoustic[50,57] and optical phonons[51,58] can be detected. For acoustic modes, the thin film behaves essentially as an acoustic cavity with a wavevector $q = n\pi/d$, where $d$ is the film thickness and $n$ is an integer. For van der Waals materials, the finite number of layers leads to the activation of interlayer vibrational phonon modes. These vibrations are associated with an out-of-plane or in-plane displacement of the layers, which are known as breathing modes and shear modes, respectively. Both types of modes are thickness-dependent and are typically used as a fingerprint to determine the number of layers ($N$) of a van der Waals material.[59–62] The thickness dependence of both types of modes can be simulated by using a one-dimensional linear atomic chain model.[44,62] This model considers each layer to be a large artificial atom with an effective mass per unit area $\mu$, connected by a spring with an effective interlayer breathing ($K_\perp$) or shear ($K_{//}$) force constant per unit area and separated by a distance $d$ given by the interlayer distance. The thickness-dependence of the breathing and shear vibrations are given by:

$$f_j^i(d) = \sqrt{\frac{K_i}{\mu \pi^2}} \sin\left(\frac{q_{N,j}d}{2}\right) \quad (1)$$

where $i$ represents shear ($i = //$) or the breathing ($i = \perp$) vibration, $q_{N,j}$ is the phonon wavevector $q_{N,j} = \pi j/(Nd)$, $j$ is the index of the acoustic mode (i.e., j= 1, 2, 3…), and $K_i$ is the interlayer force. By measuring the frequency of the breathing or shear vibration versus the number of layers, the interatomic force can be estimated by fitting the experimental data. Furthermore, the multiplication of $K_i$ by the interatomic distance provides the $C_{33}$ (breathing) and $C_{44}$ (shear) elastic constants. From eq. 1 we can also calculate the group velocity ($v_g = 2\pi \, df/dq$). In the limit as N→∞ the expression gives the limit of the cross-plane longitudinal and shear velocity, respectively.

$$v_{g_j}^i(d) = d\sqrt{\frac{K_i}{\mu}} \cos\left(\frac{\pi j}{2N}\right) \quad (2)$$

Another interesting property of these modes is that they are also sensitive to the stacking orientation and the relative intensities and peak position of the modes can vary as a function of the stacking angle.[63,64] This change in intensity or spectral position is associated with a change in the interatomic force between layers as well as a change in the crystal symmetry generated by the stacking. **Figure 1** shows how the Raman spectra changes in a low frequency Raman spectroscopy measurement of bilayer $MoS_2$ for two stacking orientations: A-A (red triangle) and A-B (grey triangles).

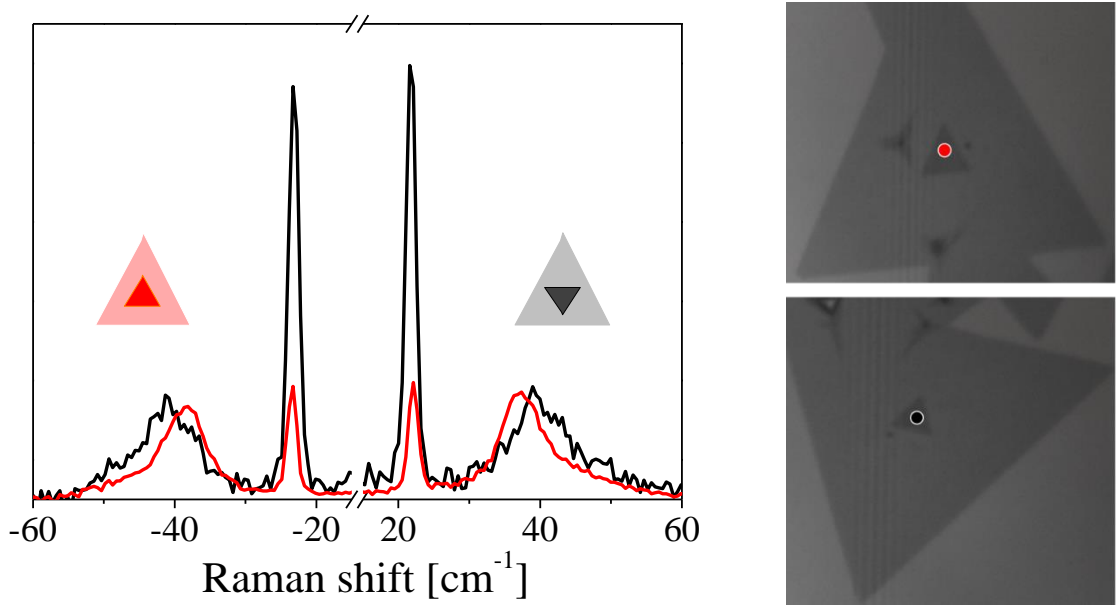

**Figure 1.** Unpolarized low frequency Raman spectra of MoS$_2$ for A-A and A-B stacking orientations. The measurement was performed using a S&I Raman spectrometer (www.s-and-i.de) using a 532 nm wavelength laser with an incident power below 100 μw focused on MoS$_2$ crystals on a SiO$_2$/Si substrate grown via chemical vapour deposition.

Beyond confinement in these types of 1D systems, two-dimensional (2D) confinement can be observed with Raman spectroscopy in self-assembled molecules (e.g., nanoporous graphene, nanoparticles embedded in a host matrix, or in phononic crystals), where phonon propagation is restricted to the in-plane direction along the direction of periodicity in the nanostructure. For the case of phononic crystals, current fabrication techniques are still limited to a minimum pitch of 10's of nm. This corresponds to acoustic confinement in the 10's of GHz (~0.3 cm$^{-1}$), which limits observation of phonons in these types of structures with Raman spectroscopy. Three-dimensional confinement has also been observed in nanoparticles, where the phonon propagation is restricted in all directions.[65]

**Laser Doppler Vibrometry**

Laser Doppler vibrometry (LDV) is a characterization technique that allows for contactless measurement of vibrational displacements by relying on the Doppler shift experienced by a laser beam when reflecting off of a moving surface. Interferometric analysis of the backscattered light combined with advanced interferometric (typically heterodyne) detection and signal analysis methods[66] are used to quantitatively reconstruct the vibration amplitude.[67] LDV finds applications at various length scales ranging from macroscopic scales (e.g. to perform contactless structural health monitoring of mechanical structures in industry applications[68,69] down to the nanoscale where it is used to characterize micro- and nano-mechanical vibrations.[70] Although LDV can be adapted to detect in-plane waves[71] or torsional motion,[72] we limit the scope here to scanning-LDV approaches as it is better suited to the study of on-chip phononic devices. In this case, the optical probing apparatus is coupled to an imaging technique that enables reconstruction of the surface displacement field (schematically illustrated in **Figure 2**).

Contactless techniques such as Raman spectroscopy and LDV are essential to sense vibrational motion without altering the local elastic properties while providing access to the natural mechanical properties. Other techniques that are discussed later on in the review can be more invasive. For example, integrated transducers such as interdigitated electrodes or plasmonic resonators introduce dissipation and can perturb the vibration frequency. Moreover, these solutions often cannot be implemented due to application-specific device geometry or material requirements. As such, interferometric optical techniques such as LDV present an advantage in that they allow for remote determination of the mechanical properties of all types of solid state systems, regardless of the material. That being said, the detection requires sufficiently high back-scattered intensity which implies that the surface reflectivity should be reasonably high. It is also worth noting that local changes of the elastic properties at the laser spot position such as changes in the mechanical frequency of a free-standing structure or changes in the reflectivity can also occur with photo-thermal absorptive materials. Standard LDV utilizes frequency-domain analysis of the mechanical motion, though time-domain studies are also possible for the elucidation of transient dynamics of mechanical motion.

Modern scanning-LDV systems fulfil the requirements to investigate nanomechanical motion in most phononic platforms. With sub-picometer displacement sensitivities,[73] these systems can be used to sense the out-of-plane displacement induced by the propagation of surface acoustic waves. Commercial scanning-LDV systems are now capable of high-frequency demodulation (i.e. above 2 GHz), which makes this technique particularly attractive for the study of acoustic platforms operating in the GHz frequency range, such as for the characterization of piezoelectric actuators[74] or for their combination with suspended nanomechanical waveguides.[75] In addition, scanning-LDV provides a singular advantage over other sensing methods by enabling displacement fields to be accurately mapped spatially, both within reasonable acquisition times and with lateral resolution limited essentially by diffraction.[76] The displacement field which is acquired point-by-point can be reconstructed over a mechanical oscillation cycle using a reference clock. This contains crucial information that is generally not obtainable with other detection techniques. For example, scanning-LDV facilitates identification of a given mechanical mode with regards to the associated finite-element simulations, which would otherwise be limited to a comparison of the mode's spectral position in experiment versus simulation. It also enables an advanced analysis of the dissipation mechanisms by identification of the loss channels, highlighting the weakness of a given design. Accounting for these loss mechanisms is particularly relevant in the context of topological phononics. Scanning the displacement field can also reveal extremely insightful information that is useful to highlight the chirality of a mechanical oscillation on both sides of a topologically protected interface.[77,78]

Despite its advantages, LDV is limited to relatively low frequencies compared to other sensing techniques such as cavity optomechanics or Brillouin light scattering. The frequency demodulation of a heterodyne signal remains limited by the laser spot size,[76] although recent progress in modal analysis has allowed for an increase in the limit demodulation frequency.[79] Furthermore, thermally excited phonon modes generally remain too weak to be accessed by LDV, especially for high frequency systems which exhibit weaker displacements. Therefore, LDV detection must often be accompanied by a separate excitation method such as interdigitated transducers[74,75] or other types of actuation.[80]

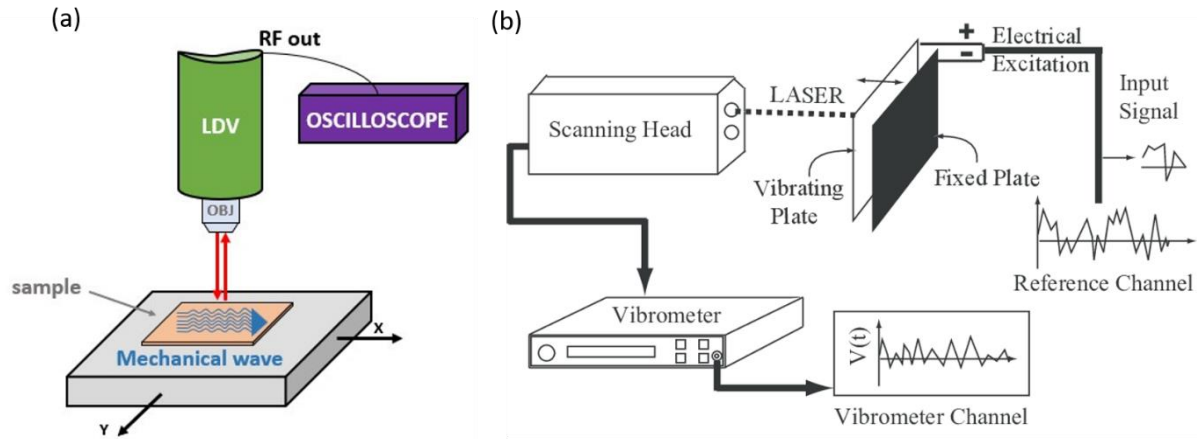

**Figure 2.** Layout of a typical LDV system. **(a)** Schematic of scanning-LDV system. **(b)** Working principle of the LDV: An electrical excitation causes the suspended sample to vibrate. The laser from the interferometer in the scanning head is focused on a sample. A photo detector records the interference of the back scattered light with the reference beam. An output voltage which is proportional to the velocity of the scanned point parallel to the measurement beam is provided by the vibrometer. Finally, the output signal is obtained as velocity or displacement signal using the velocity or the displacement decoder. **(b)** reprinted with permission from [81].

## Excitation and Detection Techniques

### Brillouin light scattering

Brillouin light scattering (BLS) spectroscopy is a non-destructive, contactless characterization technique which relies on the interaction of incident light with thermally excited acoustic phonons of a material. The incident light is inelastically scattered by phonons of the sample, modifying the frequency and wavevector of the scattered light relative to that of the incident light.[82] The mechanism is the same as that observed in Raman scattering, with the primary difference being the energy of the phonon involved in the process. Raman scattering generally probes optical phonons in the THz or sub THz range. By contrast, light in BLS is scattered by thermally excited dynamic fluctuations, or MHz-GHz acoustic phonons. These thermal excitations are governed by the equipartition principle of thermodynamics, which states that at a temperature $T$, a lattice will vibrate randomly with an average amount of energy equal to $k_B T$, where $k_B$ is the Boltzmann constant. While BLS is usually used in the frequency domain, it can also be resolved in the time domain. Time-domain BLS is a technique that allows for the generation and detection of nanometer-long acoustic pulses using ultrafast, picosecond pulsed lasers to probe the acoustical, optical, and acousto-optical parameters of materials.[83] Similarly, pumped BLS combines pulsed photoexcitation at high repetition rate with BLS spectroscopy detection.[84] This technique offers a transducer-free source and a detector of spatially confined, standing, and propagating gigahertz acoustic waves, where the photoexcited BLS signal is strongly enhanced compared to spontaneous BLS. The main mechanisms that enable the BLS process are the photoelastic[85] and moving boundary or ripple[86] effects. Both of these mechanisms are schematically illustrated in **Figure 3**. The photoelastic mechanism describes the change in the refractive index of a material caused by strain resulting from propagation of the acoustic phonons. This mechanism is dominant in transparent materials and is dependent on the penetration depth of the incident light, and consequently is independent of the incident angle of light. The ripple mechanism results from dynamic corrugations at the interface, produced by surface acoustic wave displacements. These corrugations act as a moving phase grating that diffracts light and causes a change in the energy of the scattered light. This mechanism is dominant in opaque materials and is proportional to the normal displacement of the surface acoustic wave. As a consequence the scattered light is dependent on the incident angle and independent on the material refractive index.[87] At sufficiently high powers, stimulated BLS can occur, in which the optical fields substantially contribute to the phonon population.[88] Stimulated BLS of an intense laser beam involves coherent amplification of a hypersonic lattice vibration and a scattered light wave, resulting in a strong nonlinear optical gain for the back-reflected wave. This effect is particularly relevant in optical fibers.[89]

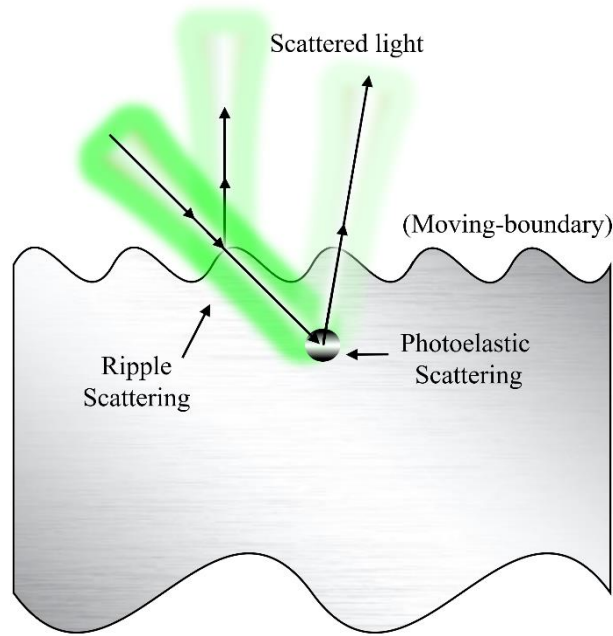

**Figure 3**. Schematic representation of the ripple and photoelastic scattering mechanisms.

As only a small fraction of light is inelastically scattered, the intensity of scattered light is approximately $10^9$ times less than that of an elastically scattered signal and is therefore difficult to detect. In Raman spectroscopy, this problem is solved by incorporating diffraction gratings. However, the several order of magnitude difference in energy between optical (~10's of THz) and acoustic phonons (GHz) around the center of the Brillouin zone greatly complicates the detection of Brillouin scattered photons. A high contrast spectrometer that is able to achieve high extinction of the elastically scattered light, is required to effectively distinguish BLS signals. This is accomplished with Fabry-Pérot interferometers (FPIs). The spectral resolution of the instrument is defined by the width of the transmitted peak, which depends on the separation of the mirrors.[90] A single FPI itself does not provide sufficient contrast to detect nonlinear scattered light. Two solutions are implemented to circumvent this problem: multi-pass systems, where the light passes multiple times through the same interferometer using prisms,[90] and a tandem interferometer configuration, where two or more interferometers are placed in series. The contrast that is obtained with two FPI in series is sufficient to detect inelastically scattered light.[91] A diagram of this configuration is shown in **Figure 4a**. The main disadvantages of such a configuration are the necessity for high spectrometer mechanical stability and the long acquisition times required for such a measurement.

To reduce the long acquisition time required using tandem FPIs, confocal microscope systems based on virtually imaged phase array (VIPA) spectrometers have been developed in recent years.[92] In these systems (**Figure 4b**), the scattered light passes through a VIPA (a solid etalon), producing a spectrally dispersed pattern in the focal plane of the spherical lens that is placed after the VIPA, and the resulting pattern is projected onto a CCD camera.[92] As is the case with a single FPI, the main limitation of a single-VIPA spectrometer is the low extinction (or spectral contrast) of the elastic signal. To improve the contrast, two VIPA spectrometers have been orthogonally placed in tandem to one another.[93] Additional strategies have been implemented to increase the contrast of this type of spectrometer, such as: placing a third VIPA spectrometer stage,[94] using a Fabry-Pérot etalon as a narrowband filter,[95] and optimizing signal collection efficiency,[96] among others. With the incorporation of VIPA based spectrometers, acquisition time can be drastically reduced down to ~100 ms, even for low-power incident light, extending the application of this technique to biological materials which are generally more sensitive to damage by light.[93] As a consequence, the capacity to measure mechanical properties non-invasively and in vivo at the cellular scale has extended applications of BLS imaging to the realm of biomedicine and biomaterials.[93–95] VIPA spectrometers with sufficient contrast to observe Brillouin signals in transparent tissues, polymers, and biological cells have already been shown.[95,97,98]

Originally, Brillouin spectroscopy was intended primarily to determine the elastic properties of bulk materials and layered structures.[99] Subsequently, it has been deployed to determine elastic properties of materials such as polymers[100] or biological systems.[101] More recently, this technique has been applied to the investigation of other physical phenomena in diverse materials and structures. One of the main applications of BLS has been the detection of confined modes in a variety of nanostructures such as nanoporous alumina,[102] thin films,[16,103] nanospheres,[31] nanowires,[104] nanocubes,[105] and core-shell structures.[106] Additionally, there has been growing interest in the use of Brillouin scattering to investigate acoustic phonons in phononic crystals[29,30,107] and to detect guided modes in phononic waveguides.[108] BLS has also been shown as a promising technique to study and detect theoretically predicted topologically protected phonon states, phonon chirality, and phonons in the hydrodynamic regime.[109]

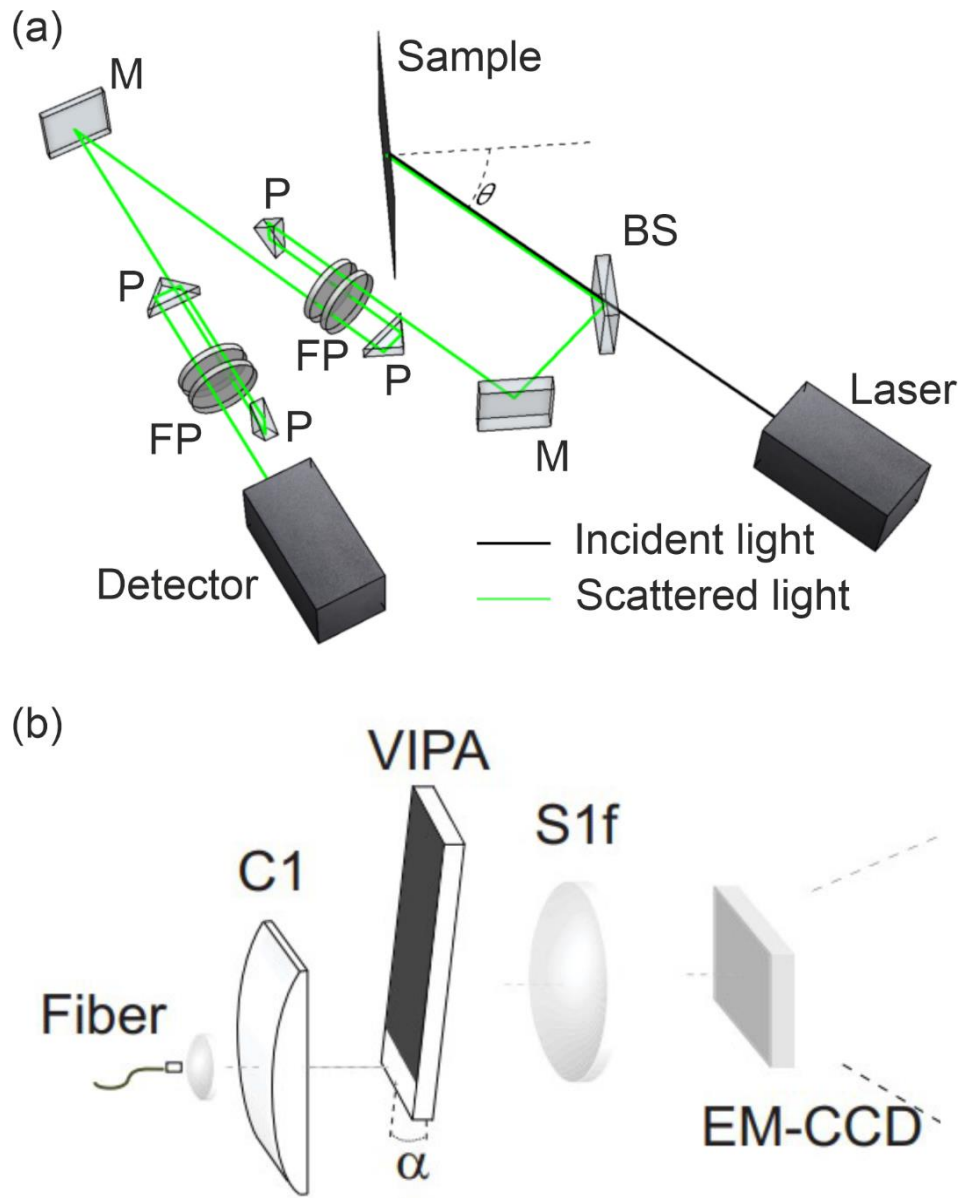

**Figure 4. (a)** Simplified scheme of a tandem Fabry-Pérot interferometer used to measure and enhance the backscattered signal. BS: beam splitter, M: mirror, P: prism, FP: Fabry-Pérot cavity. **(b)** Schematic of a single-stage VIPA spectrometer. C1: cylindrical lens, α: angle inclination of VIPA (this inclination is optional, but is sometimes used to increase the contrast of the interferometer), S1f: spherical lens. **(b)** reprinted with permission from [94].

**Scanning probe microscopy**

SAWs can be generated by applying an oscillating signal to a suitably designed set of surface gates. Acoustic methods[110] and non-invasive optical[111] and scanning probe microscopy (SPM) based techniques have been extensively employed for the detection of SAWs. However, conventional optical methods are not applicable for subsurface imaging and their spatial resolution is limited by the diffraction limit. Acoustic techniques such as scanning acoustic microscopy (SAM) provide improved spatial resolution as they use ultrasonic waves with higher frequencies, though the spatial resolution is still limited by the Rayleigh criterion. These techniques have been applied for detecting cracks, voids, and delaminations within micro-devices.[112,113] Commercial SAM setups usually operate at frequencies between 10 and 100 MHz.

To overcome the diffraction limited resolution in acoustic microscopy, near-field scanning probe techniques have been developed such as scanning near-field ultrasonic holography (SNFUH).[114] SNFUH uses ultrasonic plane waves that enter through the bottom of a sample for the characterization of the elastic properties of surfaces as well as of buried structures (or material) in 3D geometries,[115] providing a quantitative method for surface acoustic wave detection. The technique uses an aperture or tip as an antenna, which is brought into close proximity with the sample. The spatial resolution is then determined by the size of the contact area between the antenna and sample. One of the earlier demonstrations of this concept was by Zieniuk and Latuszekref,[116] who built a scanning near-field acoustic microscope using a pin probe as the antenna, although this was limited to a poor spatial resolution of about 10 μm. Progress in SPM techniques has since yielded higher resolution near-field acoustic imaging. Today, SPM methods incorporate nanometer size tips (~10 nm), which allow for nanoscale near-field acoustical imaging.

Several works have reported acoustic imaging instruments based on scanning tunneling microscopy (STM).[117–119] In these methods, the surface displacement is monitored either through the tunneling current or through the ultrasonic transmission through the tips. Following this, Chilla et al. used ultrasonic STM to detect both the out-of-plane and in-plane components of surface acoustic waves.[120] While STM presents many advantages, its application is limited only to conductive samples. As an alternative, atomic force microscopy (AFM) has no such limitation which resulted in significant efforts to realize acoustic imaging with AFM. Various different methods for detecting ultrasonic vibrations with an AFM were proposed. First, Yamanaka et al.[121] developed an AFM-based method to detect ultrasonic vibrations of the sample at frequencies much greater than those of the AFM cantilever resonances. Later, the same group used an ultrasonic force microscope (UFM) for imaging subsurface defects in graphite with a resolution better than 10 nm.[121,122] To date, this technique has been used for nanoscale imaging of structural and mechanical properties of complex nanostructures and thin films.[123] Similar ultrasonic AFMs were used for the mechanical mapping of material surfaces.[124–126] Furthermore, the detection of ultrasonic surface vibrations with MHz range bandwidth has been accomplished by adding an optical knife-edge detector in an AFM setup.[127,128] In addition, AFM based probe systems for the detection of laser-induced SAWs have been reported.[129] One of these systems is shown in **Figure 5a**. A pulsed laser beam irradiates a surface at point A and induces a SAW which propagates along the surface. The vibration of the surface at point B is then measured by an AFM probe.

To improve detection sensitivity, the AFM technique must be operated at the resonance frequency of the cantilever in contact with the sample. Consequently, several ultrasonic AFM techniques have been developed based on different excitation and detection schemes for ultrasonic waves such as ultrasonic atomic force microscopy (UAFM)[130] and atomic force acoustic microscopy (AFAM).[131,132] The latter uses ultrasonic waves between the AFM scanner and a sample. The sample is bonded to an external transducer which generates longitudinal ultrasonic waves with a center frequency of several MHz supplied by a function generator. The ultrasonic waves propagate through the sample resulting in out-of-plane vibrations of the sample surface. Those vibrations are monitored via deflections of the AFM cantilever in contact mode operation. To simultaneously extract surface topography along with the acoustic amplitude and phase of the cantilever vibrations, the signals are analyzed with a lock-in amplifier. A schematic of an UAFM technique is shown in **Figure 5b**.

Further progress has been achieved with the development of ultrasonic near-field optical microscopes (UNOM) that enable local mapping of ultrasound with deep sub-optical wavelength spatial resolution. As an example, Ahn et al. generated ultrasonic waves with a pulsed laser and detected them with a scanning near-field optical probe over a broad frequency bandwidth (see **Figure 5c**).[133] They used a plasmonic probe which enhanced the scattering of evanescent light at the probe-tip and enabled reliable measurement of the motion of the surface. Lastly, an AFM-based technique has recently been developed for the imaging of hyperbolic phonon polaritons in two-dimensional (2D) materials,[134] demonstrating the potential of SPM methods for characterizing nanomaterials.

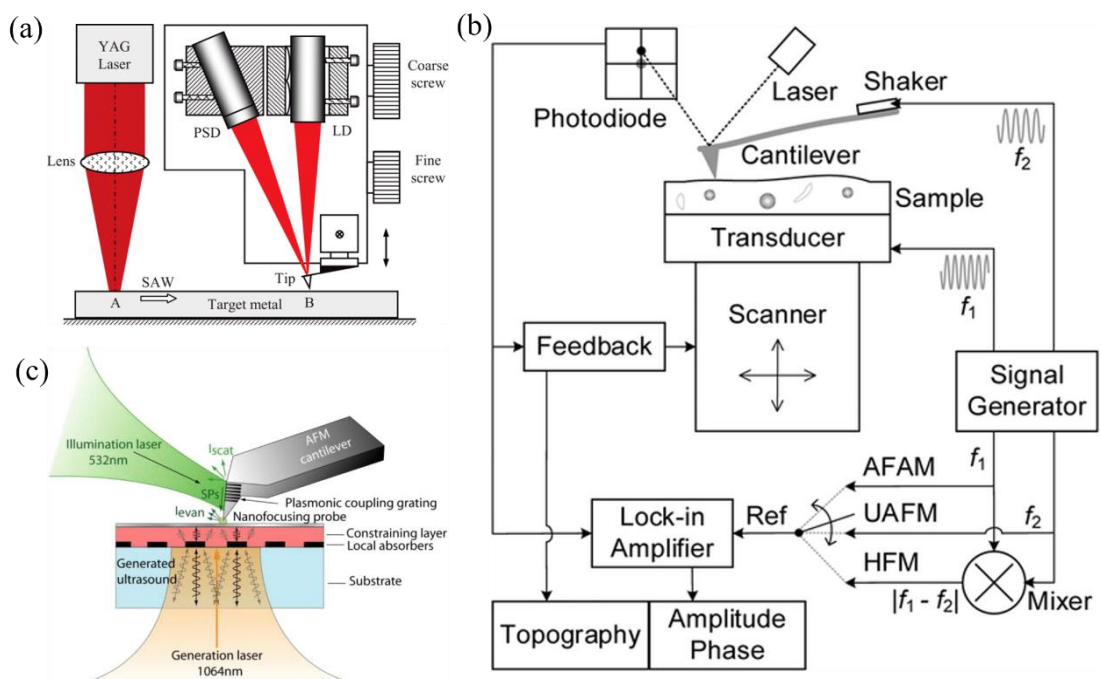

**Figure 5.** **(a)** Scheme of an AFM-based probe system for detecting laser induced surface acoustic waves. **(b)** Schematic of an AFAM setup. The vibrations of the cantilever are excited either by a transducer below the sample (transducer 1) or by a transducer which is positioned on the top end of the cantilever (transducer 2). The low frequency components of the beam deflection signal are used to control the static deflection of the cantilever. **(c)** Schematic of an ultrasonic near-field optical microscopy setup using a plasmonic probe. **(a)** reprinted with permission from [129], **(b)** with permission from [135], and **(c)** with permission from [133].

**Cavity optomechanics**

Cavity optomechanics (OM) studies the interaction between light and mechanical motion. Within cavity OM systems, a driving laser couples to mechanical motion via the radiation pressure force. The mechanical motion also modulates the output optical field, which is measured. This interaction is enhanced within an optical resonator. The mechanical motion subsequently also affects the optical field. In this manner, cavity OM allows for the control and detection of nanomechanical motion using light. The canonical system for cavity OM is that of an optical cavity created between two mirrors, where one mirror is movable by suspension or by being attached to a spring. There have been many additional examples of OM systems at various length scales to date, and a more general treatment of OM can be found in other reviews.[37,136,137] While the entire field of OM is quite broad, and can in the most general sense be used to describe any system in which mechanics are probed by light, we limit the scope of this section of the review to the detection of mechanical modes solely in on-chip cavity OM crystal systems at the micro/nano-scale, while many of the same ideas are further applicable to cavity OM systems at other length scales as well. Many other exotic systems also exist such as optical tweezers and levitated nanoparticles[138–142] or levitated superfluids.[143–146]

Realizing cavity OM at the nanoscale can be difficult, and common systems towards this end include microwave circuits,[147] micro/nano-mechanical resonators,[148,149] and OM cavities in 1D (e.g. nanobeams)[150–158] and 2D (photonic/phononic crystals).[159–161] At these scales, these systems sustain mechanical frequencies from the kHz up to the 10's of GHz scale. Recently, some systems have demonstrated resonant mechanical frequencies up to 100 GHz.[162] Limitations in fabrication resolution at the nanoscale and detection of these high frequency oscillations prevent the realization of nanoscale OM systems at even greater frequencies. Cavity OM systems enable extremely sensitive readout of mechanical motion, even when the motion is driven by thermal forces. However, the detection of such weak perturbations of the optical field requires the use of appropriate tools. Radiofrequency (RF) photodetectors and electrical spectrum analyzers (ESAs) constitute a common approach to extract this signal, although low-noise amplification of the optical and/or the electrical signal can sometimes be essential to extract these signals from the residual noise of the detectors (e.g. dark noise/shot noise). Interferometric methods, such as homodyne[163] or heterodyne detection[164,165] can also be implemented to subtract the noise originating from laser fluctuations. Measurements are often frequency-resolved, although time-resolved OM measurements can also be done to study the dynamics of a system. These time-resolved measurements tend to be limited to relatively lower frequency mechanical modes. While the frequency resolution that can be obtained is usually limited by the resolution bandwidth (RBW) of an ESA, this is generally not a limiting factor in cavity OM, as OM resonances are at least on the order of a few 10s of kHz in the lasing regime, or on the order of several MHz in the thermal regime. These systems tend to be limited by the displacement sensitivity (smallest displacement change that can be detected) and the dynamic range (measurement limit due to nonlinearities in the cavity system or electronic limitations of the detection equipment) which compete with one another.[166] The displacement sensitivity is limited by the quality factors of the optical resonances within the cavity, and consequently also require extremely narrow-linewidth, low noise lasers. Optical quality factors on the order of $10^5$-$10^6$ have been reported in nanoscale cavity OM platforms at room temperature, though much higher quality factors can be reached at cryogenic temperatures.[167] Within these systems, displacement sensitivities less than $fm/Hz^{1/2}$ can be reached, although this limits the dynamic range to a few nanometers.[168] The dynamic range is limiting since any mechanical motion of the cavity is detected by tracking the linear part of an optical resonance. The consequence of this is that relatively low optical powers must be used, such that a linearized OM approximation can be used, and any non-linearities (e.g., thermo-optic, free carrier dispersion, or Kerr effects) that affect the linear regime of a cavity optical resonance are avoided. Noise limitations such as dark noise, a generated current due to thermal excitations that is present even in the absence of incident photons, and shot noise, an electronic noise resulting from the discrete flow of charged particles, in any detection readout electronics such as the ESA and photodetectors also must be considered. Even the dynamical back-action is itself a source of noise (radiation pressure shot noise).[169] Many of these limitations are not unique only to cavity OM, though they become practically relevant due to the much better displacement sensitivities that can be achieved in cavity OM which are generally much smaller than many other experimental techniques.

OM systems are often engineered to be monomode photonic crystal cavities. Coupling into these guided optical modes of an OM cavity can be complex, often requiring the use of a bus waveguide, grating couplers, or a tapered fiber that enables coupling from free space. Using a tapered fiber from which light can evanescently couple to a cavity enables phonon detection at high spatial resolutions as light can be coupled even to specific individual nanocavities. Recent work has studied the ability to access and excite mechanics by driving with Anderson-localized optical modes,[161,170] while direct experimental evidence of localized mechanical modes still has yet to be demonstrated. Spatial mapping of the mechanics is also possible, following a procedure such as that demonstrated by Ren et al.[171] in which a 300 MHz topological channel can be probed by optomechanically interrogating the optical cavities embedded in a lower length scale photonic crystal. In this example, the dissipation occurring at the sharp corners of the acoustic channel allow for a comparison between trivial and topological waveguides.

While OM serves as a powerful tool to detect and control phonons in the kHz-GHz regimes at the micro/nanoscale, OM ideally requires strong confinement (i.e. high quality factors) of both the photons and of the phonons, to increase the quality factors of the optical and mechanical modes while maintaining a small mode volume to enhance photon-phonon interaction and coupling. Within these types of systems, careful consideration of design is required to enable this confinement and utilize OM detection. Fabrication imperfections can lead to the breaking of symmetries and allow coupling between the different phononic symmetries, though this can be overcome by surrounding the OM crystal with a phononic radiation shield that possesses a full phononic gap.[151,153] These careful design considerations limit the use of OM techniques for phonon detection to those in which photonic/phononic cavities can be incorporated. However, OM crystals, particularly those based on silicon platforms, are readily incorporated into integrated photonic/phononic circuits due to their compatibility with existing CMOS technologies.

**Pump-probe**

The invention and widespread adoption of femtosecond (fs) lasers opened the possibility to study non-equilibrium systems with fs resolution dynamics. When a material is illuminated by an ultrashort pulse of light, rapid dynamics are triggered. A photon-electron interaction will occur, followed by thermalization with the ion lattice. This process converts optical energy into mechanical energy, which produces a photoinduced stress. The relaxation of this stress can launch coherent mechanical vibrations (i.e., coherent acoustic phonons) in a system with frequencies up to 100's of THz. The different modes and the frequencies at which a system vibrates at depend on its shape, composition, and surrounding medium.[172–177] Generation and characterization of phonons and measurement of the different oscillation mode frequencies is usually done with pump-probe techniques. The technique generally begins with a single mode-locked laser that is divided into two separate beams. One is the excitation beam (pump) while the other is the detection beam (probe) with which the state of the sample at any given moment can be detected. During a measurement, they are temporally and gradually delayed from one another, and the transmittance/reflection of the probe is collected. This process is schematically illustrated along with a typical expected optical response in **Figure 6**.

Following the incidence of the pump beam and the subsequent conversion from optical to mechanical energy, a strain is generated. The dielectric properties of the sample around the illuminated area are modified momentarily, which causes a change in the refractive index that depends on the excited modes. The affected area of the sample then oscillates at a certain frequency, until it once again reaches thermal equilibrium. The probe beam allows for detection of these vibrations. The oscillatory modification of the dielectric properties will cause a change in the transmission/reflection of the probe beam in the time domain.[176] From the collection of the probe beam, the state of a system can be determined at any given moment in time, with each moment determined by the time delay between pump and probe. The dynamics launched by the pump beam can be readily reproduced. The time delay is typically generated after sending the probe beam through a linear translation stage controlled with a stepper motor. For each step (in units of distance) of the stage, there will be a temporal separation between the pulses that can be calculated after every step by considering the light's velocity. The temporal resolution of the measurement depends on the width of the laser pulses, usually on the order of 10's of fs. Since the function of the probe is to interrogate the sample, ideally without affecting it in any manner, the power of the probe tends to be an order of magnitude smaller than the pump, such that the probe beam pulse ideally does not modify the state of the system. The polarization and wavelength of the pump, the probe, or both, can be readily manipulated, as required by the specific phenomena of interest. For example, a non-linear crystal is commonly placed in the path of one of the beams to modify its wavelength. This allows the beams to be easily separated from one another at the detector using color or interference filters.

The pump-probe technique is versatile with wide-reaching applications and has been used to measure all types of matter from gases and liquids[178,179] to solids.[172,180,181] Many diverse systems have been characterized such as metallic films,[174] acoustic nanocavities,[182] piezoelectric materials,[183] and 2D materials.[173,184,185] Notably of interest here, this technique has been applied to measure coherent phonon modes and their lifetimes over a large number of frequency regimes. A phonon's lifetime refers to the timescale over which a phonon is scattered or attenuated. The reason for this attenuation can be due to either its collision with impurities or defects within a sample or its boundaries (extrinsic mechanism), or due to the intrinsic anharmonicity of the lattice, which occurs even in perfect crystals (intrinsic mechanism). Together with the group velocity, the phonon lifetime defines how far a phonon can carry its energy, which is known as the phonon mean free path. Despite the fundamental importance of this parameter, accurate measurements of phonon lifetimes are challenging, and their values are unknown in most materials. Although silicon is the most important material for nanoscale devices, there are very few direct measurements of phonon lifetimes in the gigahertz to terahertz range even in this platform.[18,186–189] For bulk samples, the phonon lifetimes tend to be limited primarily by anharmonic interactions (phonon-phonon scattering) and by impurity scattering, though to a lesser extent. However, as the length scale of a system is reduced, the phonon lifetime becomes limited primarily by surface roughness. One state-of-art technique of generation and detection of coherent phonons in nanostructures is the Asynchronous Optical Sampling spectroscopy (ASOPS).[190] The ASOPS method is based on traditional ultra-fast pump-probe techniques, though it produces pulses from two mode-locked femtosecond titanium-sapphire lasers with slightly detuned repetition rates rather than incorporating a mechanical delay line for the temporal dephasing of the lasers. ASOPS uses the detuning of the repetition rate between the pump and probe pulses. This detuning creates a monotonic temporal window that allows the dynamics of the system under study to be scanned without requiring any readjustments of the mechanical stage. The scan rate is determined by the difference in frequency, $\Delta f_R$, and the temporal window is given by the inverse of this difference, $1/\Delta f_R$. For example, if the repetition rates of the pump and probe are 1 GHz and 0.999999 GHz, respectively, a temporal window of one nanosecond can be probed in 0.1 ms.

The first measurements involving the generation and detection of acoustic phonons by picosecond laser pulses were performed in the 1980s.[191,192] The technique was later applied to measure the attenuation of phonons in amorphous $SiO_2$, for frequencies up to 440 GHz.[193] Pioneering experimental studies of the detection of confined phonons using a pump-probe technique were performed by Thomsen et al. in 1984.[191] They used the modulation of the optical transmission through picosecond pump-and-probe to detect coherent phonons in α-$As_2Te_3$ and cis-polyacetylene thin films. Later, Thomsen et al. used the reflected signal to detect confined phonons in α-As2Te3, α-Ge, α-As2Se3, and Ni films.[192] Today, this method is widely used to detect confined phonons in many types of nanostructures,[176] van der Waals materials,[185] and topological superlattices.[182] For example, Arregui et al. studied the generation of coherent acoustic phonons with frequencies on the order of 100's GHz in topological nanocavities,[182] Lin K. et al., demonstrated the generation of acoustic pulses injecting light carriers in piezoelectric materials,[183] and Lee et al. and Miao et al. studied the generation and detection of coherent acoustic phonons in Black phosphorus.[173,184] A more exhaustive general analysis of the systems in which coherent acoustic phonons have been studied using the pump-probe technique can be found in Ruello P. *et al.*[176] and Vialla F. et al.[185]

The generation of coherent acoustic phonons also have been studied in a wide range of different plasmonic nanostructures, which we discuss here as an example of the application of the pump-probe technique. The field of nanoplasmonics is a rapidly developing field concerned with the study and application of electron oscillations at the metal-dielectric interface of metal nanostructures which has seen widespread interest due to unique properties such as nanoscale light confinement and geometrically tunable plasmon resonances.[194–199] These unique properties often enable extremely strong field enhancements due to the confinement of light down to subwavelength volumes. These unique properties have opened many new avenues for research, extending all the way to applications in phonon generation. Plasmonic nanometer-sized structures, or plasmonic nanoantennas, have been shown to be highly efficient for the generation of coherent acoustic phonons after being excited by an ultrashort light pulse.[200] After the excitation of a plasmonic nanoantenna with pulsed light at the appropriate wavelength (generally its dipole plasmonic resonance, or interband transition), an excited electron population is produced in the metal, followed by thermalization and heating of the lattice. The thermal equilibration of the hot electrons with the lattice ions launch mechanical vibrations (i.e., coherent acoustic phonons) at the frequencies of the eigenmodes of the structure.[175,201] These mechanical oscillations in turn modulate the optical response of the nanoantenna, which is detected in an experimental pump-probe configuration. Plasmonic structures have the peculiarity that the optical modulation signal can be enhanced if they are excited and/or detected in the surface plasmon resonant regime. This process incorporates nanoantennas as local mechanical nano-resonators with tuneable frequencies that depend on the size, shape, and composition of the antenna, in addition to the mechanical boundary conditions, with frequencies ranging from a few GHz up to the THz regime.[201–203] For example, Della Picca F. et al.,[175] demonstrated frequency tunability via the imposition of a physical restriction by adding silica patches on gold nano-rods to modify the frequencies of the normal oscillation modes using a degenerate pump-probe configuration at the plasmonic resonance of the nanoantennas. They detected frequency changes relative to a "naked" rod (i.e., without silica patch additions) based on the specific placement and geometry of these patches. Berte R. et al.,[204] extended this idea, measuring the frequencies of coherent acoustic phonons and showing how these light induced vibrations in both gold nano-rod and nano-disc nanoantennas allow surface acoustic waves to be launched through the substrate at the frequencies of the excited modes, with size- and geometry-dependent oscillation frequencies. Here, a non-degenerate pump-probe configuration was used, with the pump beam tuned at the interband transition of the gold (~400 nm wavelength) and the probe near the plasmonic resonance of the nanoantennas (~800 nm wavelength). In the same vein, Boggiano H. et al.,[205] introduced an optical method for nanoscale measurements of the mechanical moduli of polymers at GHz frequencies based on light induced coherent acoustic phonons using gold nanorods. They analyzed the frequency change when the antenna is surrounded by air versus when it is surrounded by PMMA, which enables determination of the mechanical properties of the PMMA. They also excited the sample in an interband transition of the gold nano-rods and tuned the probe near the plasmonic resonance to enhance the detection sensitivity.

Phonon generation via the pump-probe technique with plasmonics is not limited only to gold nanoantennas. For example, Ostovar B. et al.,[206] characterized acoustic vibrations on the order of 10's of GHz, as well as the decay times and Q factors, induced on colloidal aluminium nanocrystals particles of different sizes and shapes using the pump-probe technique. Furthermore, the generation of coherent acoustic phonons is not exclusive only to nanoparticles (where a nanoparticle is defined in this context as an antenna with dimensions less than or comparable to the incident wavelength of light). Larger plasmonic nanowires have also been used, excited by impinging on them with an elliptical pump and causing them to emit acoustic waves at their eigenfrequencies, due to their coupling with the substrate. This was shown by Imade Y. et al.,[203] in the GHz regime where they study the generation of acoustic phonons in plasmonic nanowires and the propagation of the different modes through the substrate in surface acoustic wave form. Depending on the geometry of the plasmonic structures, the frequency and directionality of the surface acoustic waves emitted can be controlled after the relaxation of the phonons.[203,207] These examples demonstrate how coherent acoustic phonons can be induced and how their oscillation amplitude, frequency, and vibrational modes can be controlled. They further demonstrate how these phonons can be used to characterize systems in different manners, such as the mechanical properties of a polymer or the imperfections of a substrate as acoustic waves are sent through it.

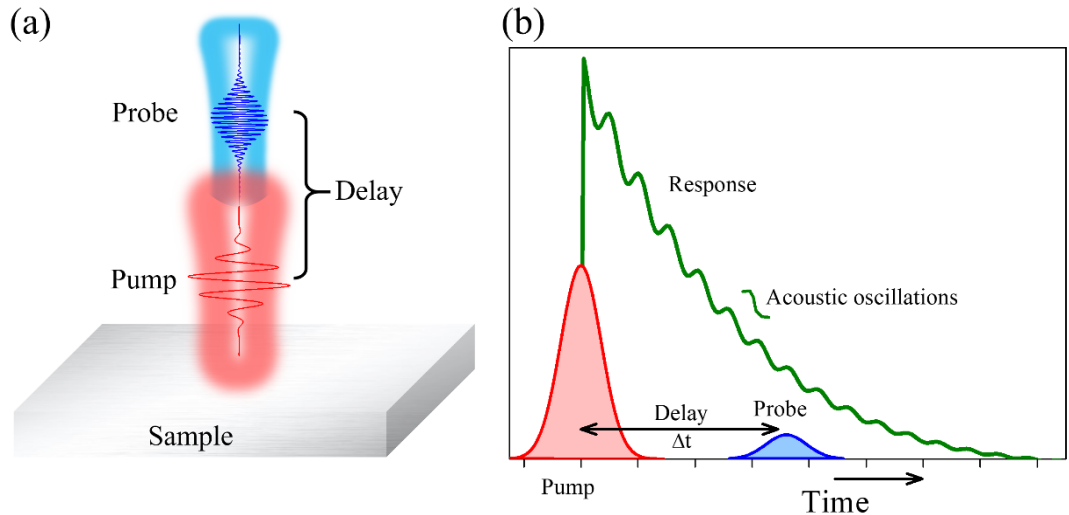

**Figure 6.** Pump-probe technique. **(a)** Pump-probe measurement configuration. **(b)** Typical optical response of a sample under pump-probe excitation.

**Interdigitated transducers**

Phonons readily interact with a variety of other quanta, fields, and forces, enabling the realization of devices based on phonon-mediated modulation. However, such devices rely on the emission and detection of traveling phonons, which in most cases tend to require bulky experimental setups. To overcome this challenge and leverage these industrially relevant emerging technologies, advanced methods that combine platforms have been recently proposed. These platforms are mainly based on optomechanical interactions, to exploit their technological and functional complementarity. This section provides a brief overview of other remarkable approaches that take advantage of interdigitated electrodes or transducers (IDTs) and surface acoustic waves (SAWs) that lay the foundations for a platform for integrated phononics and photonics.

The evolution of IDTs has been constant since their first demonstration in the 60s.[208,209] IDTs are metallic electrode arrays that were traditionally fabricated on a piezoelectric substrate. In this configuration, the piezoelectric properties of the substrate enable electromechanical coupling (i.e., actuation), where a sinusoidal electrical signal on the order of the IDT array period that is applied to the

IDT generates a SAW, a surface-localized guided mechanical wave that propagates along the plane of a surface (see **Figure 7a**). Conversely, IDTs can detect a propagating SAW by converting the mechanical wave into an electrical signal. The potential realization of many types of physical sensors using IDTs has led to an explosion of their use in modern mainstream technologies even many decades following their inception.[208,210–213] In more recent years, improved nanofabrication techniques have resulted in the development of high frequency IDTs operating up to the GHz regime.[214,215] Fundamental research has exploited the ability to generate phonons from a RF signal as well as read them out with excellent sensitivities,[216–218] and has used them to interact with a wide range of different systems such as optomechanical systems,[216] nitrogen-vacancy centers,[217] and superconducting qubits.[219–221] The wide variety of IDT designs (see **Figure 7**) make the category of IDTS very peculiar when compared to the other techniques that are discussed within this review. Generally, IDTs operate from the kHz regime up to ~ 10's of GHz.[222] The exact characteristics (e.g., frequency and spatial resolution, frequency range, bandwidth, signal shape, etc.) that can be obtained are highly dependent on the specific IDT design. For example, bandwidth is dependent on the number of digits in the IDT, where more digits result in a more intense signal but a narrower bandwidth. The frequency resolution is a practical limitation, dependent on the experimental equipment available. There are additional considerations when incorporating IDTs that may limit their application, such as the invasiveness of their integration into the underlying sample.

While the field of SAWs generated via IDTs in piezoelectric materials is quite a developed field that has resulted in a plethora of applications,[208,211] the electromechanical coupling that is dependent on the requirement of a piezoelectric material substrate can be a limiting factor in many applications. Notably, the lack of CMOS compatibility impedes monolithic integration with electronics. These ideas have already been extended to the realm of optomechanics, where existing platforms in standard silicon-on-insulator (SOI) have demonstrated simultaneous support of optical and mechanical signals with the goal of analog optomechanical signal processing on a single chip. Recently, Munk et al. proposed a surface acoustic wave-photonic device fabricated on SOI, completely foregoing a piezoelectric substrate, and operating at frequencies up to 8 GHz (**Figure 7b**).[223] The photonic device uses surface waves that are launched through the thermal absorption of a modulated optical pump pulse (1540 nm) in a gold grating

IDT. The resulting strain from thermoelastic expansion and contraction of the grating is transferred to the underlying silicon device layer, which in turn, can excite a surface acoustic mode of the SOI layer stack. Detection of the surface acoustic waves is based on the photo-elastic modulation of an optical probe in standard race-track resonators. They demonstrate that this photonic device can serve as a discrete-time microwave-photonic filter based on acoustics, with high relevance to signal processing applications. Based on a photonic–phononic emit–receive process, Kittlaus et al.[224] proposed a RF filter that uses high-Q phononic signal processing. The architecture is based on an all-silicon photonic-phononic emitter-receiver that exhibits high-fidelity narrow-band filtering (5 MHz) and has the advantage of using a single intensity modulator to encode RF signals onto light. Such a device converts intensity-modulated light in the emit waveguide into phase-modulated light (probe) in the receive waveguide. This is achieved through a linear acousto-optic scattering process mediated by a resonant Lamb-like acoustic mode within a silicon membrane. The probe light signal is filtered through the phononic response of the device (via the acoustic wave generated by the pump) and encoded onto the probe wave by the resulting phase modulation. This device exhibits record-high modulation efficiency compared to state-of-the-art phononic emit-receive devices and exhibits robust performance as a RF-photonic filter.

Other more complex electro- and piezo-optomechanical platforms could help to further enable integrated photonics and phononics on a chip. The following examples are potentially straightforward to integrate onto a chip, though they also support electrical actuation, in addition to optical and mechanical signals. Recently, a proof-of-principle technology platform was proposed by Navarro-Urrios et al (**Figure 7c**).[225] This platform consists of a nanoelectro-opto-mechanical system that supports the coexistence of electrical, mechanical, and optical signals on a chip. Aluminum IDTs on an aluminium nitride layer generate coherent mechanical waves (2 GHz) that interact with the optical fields confined in nanocrystaline silicon optomechanical nanobeam cavities. The surface acoustic waves generated by the IDTs are converted into guided mechanical waves supported by the nanobeam. At cavity resonance, these mechanical waves strongly interact with the confined optical waves in the released nanobeam OM cavity, enabling microwave radiofrequency-to-optical conversion at room temperature, with a peak sensitivity below 3 phonons. Mayor et al., recently proposed a mechanical waveguide of LiNbO$_3$ (LN) on sapphire (LISA) in an attempt to combine electrical, mechanical, and optical signals altogether onto a single platform, to achieve efficient electromechanical wave transduction for phononic circuits.[226] This waveguide combines the high piezoelectric coupling coefficient of LN with the confinement of the mechanical wave achieved through index guiding, due to the slower propagation in LN relative to that in sapphire. The higher refractive index of LN (≈2.2) compared to sapphire (≈1.7) also allows for simultaneous optical guiding within the same waveguide. The evanescent mode profile in the sapphire substrate enables coupling between adjacent waveguides. With this platform, mechanical waves at ≈3.23 GHz are efficiently excited by Al IDTs and guided through the LN ridge waveguide, at room temperature. Delay lines, racetrack resonators, and meander line waveguides are all demonstrated in the proposed platform. Furthermore, this platform has the advantage of being "unreleased" in that the device layer is not suspended, which facilitates its fabrication, scalability, and integration with other photonic circuits. The lack of suspension allows for more ready incorporation with existing CMOS technologies and greatly simplifies the fabrication of dense monolithic all-in-one photonic and phononic circuits.[226,227]

These advances in complex electro- and piezo-optomechanical circuits show promising avenues for optimal information transmission using multi-state variables in a single chip. Beyond this, they can find applications in integrated microwave photonics, microwave quantum technologies at cryogenic temperatures, and in general, quantum and classical phononic circuits and systems. The rich and diverse physical phenomena that manifest in these electro- and piezo-optomechanical devices may draw a new technological frontier with capabilities beyond the reach of current technologies.

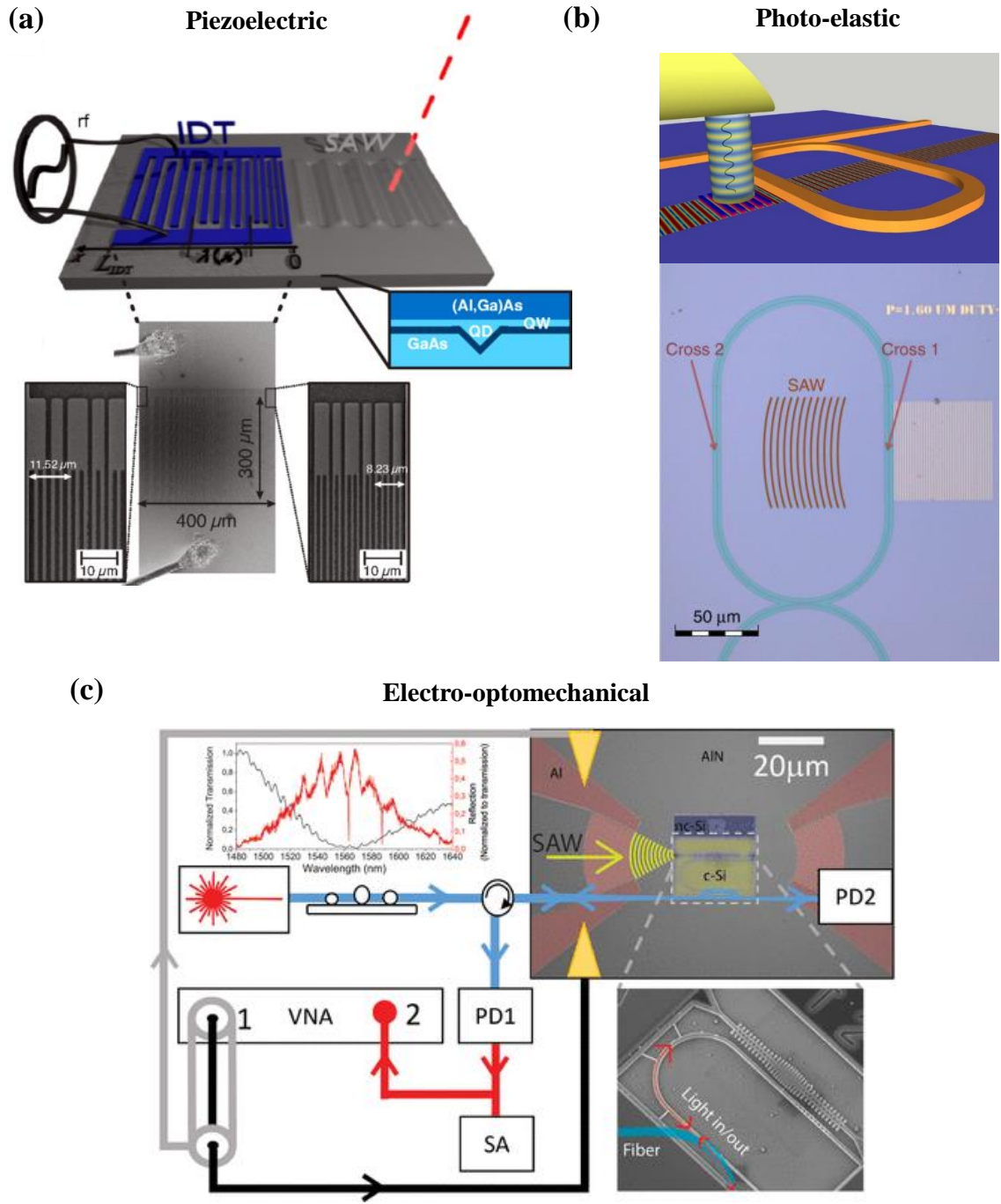

**Figure 7.** Examples of IDT systems based on: **(a)** Piezoelectric actuation of acoustic waves; device with metal IDT patterned onto an (Al,Ga)As-based heterostructure (top); SEM images of the IDT (bottom). **(b)** Photo-elastic coupling; schematic of device where modulated light is absorbed and converted into acoustic waves via thermal expansion of illuminated gold grating (top); optical microscope image of a race-track resonator waveguide and gold grating (bottom). **(c)** Electro-optomechanical coupling; experimental setup of the nano-electro-optomechanical platform where Al

concentric IDTs are used to piezoelectrically launch phonons in a nc-Si optomechanical system which is embedded in photonic circuitry (SEM image, bottom right). (a) reprinted with permission from [228] and (b) with permission from [223]. (c) reprinted with permission from [225]. Copyright 2022 American Chemical Society.

**Summary/Conclusions**

The field of nanophononics is an active and growing field of research, as demonstrated by the many examples provided here. Phonons themselves play a fundamental role throughout various scientific processes, and a complete understanding of their mechanisms and effects is required to realize next generation applications. With the recent advancements in nanofabrication technologies, the limiting factor in further advancement of the field of phononics is largely practical, due to challenges associated with the excitation and detection of phonons or vibrations. Here, we focused on highlighting six of the available techniques for manipulating acoustic phonons, and provided an overview of the advantages, disadvantages, considerations, and general development of these techniques, which are summarized in Table 1. Cavity OM and IDTs are a particular case among these techniques, in that they require specific nano-architecture on the substrate to enable their incorporation, unlike the other techniques which allow for direct measurement.

**Table 1.** Comparison of experimental characteristics of each experimental technique discussed in this review.

| Experimental Technique | Frequency Range | Frequency Resolution | Spatial Resolution | Time/Frequency Resolved | Spatial Mapping | Additional Considerations |
|---|---|---|---|---|---|---|
| Raman | 3 – 4000 cm$^{-1}$ (0.089 – 119.9 THz) | 0.5 cm$^{-1}$ (14.98 GHz)* | Optically limited ($\lambda/2$) | Yes/Yes | Yes | Requires Raman active mode |
| LDV | DC (kHz) – 2.5 GHz | Few kHz | Optically limited ($\lambda/2$) | Yes/Yes | Yes | Detection of out-of-plane motion |
| BLS | 0.1 – 1200 GHz | O (MHz) | Optically limited ($\lambda/2$) | Yes/Yes | Yes | Long acquisition times |
| SPM | 10 kHz - 100 MHz | A few kHz | < 10 nm | Yes (~µsec)/Yes | Yes | Differentiate the contribution to |
| | | | | | | contrast by surface versus sub-surface material properties |
| Cavity OM | kHz – 10's of GHz | 0 (kHz)** | - | Yes/Yes | No | Designs limited to sufficiently strong OM cavities |
| Pump-Probe | GHz – THz | 10's of MHz | Optically limited ($\lambda/2$) | Yes/No | Yes | - |
| IDTs | kHz – 10's of GHz | Design-dependent*** | - | Yes/Yes | No | Design-dependent; Invasive; Requires piezoelectric material; Can also generate acoustic waves |

*Dependent on laser wavelength, spectrometer focal length, and grooves/mm in the grating.

**OM systems are generally limited not by frequency resolution, but rather by displacement sensitivity and dynamic range which compete with one another

***Frequency resolution depends on the experimental equipment available. Bandwidth is set by number of IDT digits

**Emerging directions: Dirac-acoustic materials**

Since the pioneering work on the engineering of the acoustic band structure through a two-dimensional periodic array,[229–231] PnCs have become a rich and emerging research field. The precise tuning of the dispersion relation has allowed for the observation of exotic phenomena far beyond those found in nature. For example, Dirac cones can be generated by simply tuning lattice symmetry and fill factor.[232–235] The Dirac cone manifests as a crossed linear dispersion relation at a point in reciprocal space and is highly robust even under disorder and perturbations. Dirac materials are described by a massless Dirac-like equation and their discovery is considered to be one of the great achievements of condensed matter in the last century. Along these lines, acoustic or phonon topology has recently attracted significant scientific attention. In contrast with topological electronic materials, acoustic topology does not depend on the atomic arrangement of materials. Rather, its properties and manipulation depend only on geometrical parameters, greatly reducing the difficulty associated with the fabrication of structures that exhibit phononic topology.

Xia et al.[235] showed that small variations of the geometric parameters can shift the position of the Dirac point without opening it. A break in the mirror symmetry of the geometry can open the dispersion at the Dirac point, yielding directional or complete acoustic band gaps. Lu et al. predicted[236] and later demonstrated[237] the creation of acoustic vortices in a 2D waveguide via rotational symmetry of a crystal. Using a hexagonal array of triangular rods, Lu demonstrated the existence of an acoustic analogue of the quantum valley-Hall effect. Complementary observations were reported by Yang et al.[238] adopting a similar structure, where an array of triangles was fabricated on a silicon chip using CMOS microfabrication techniques. Other unusual properties have also been observed by concatenating two or more lattices with different symmetry, e.g., acoustic pseudospin,[239–242] and by a smooth breaking of the symmetry such as is in the case of acoustic pseudomagnetic fields.[243]

Until now, efforts in topological phononics have generally been limited to low-frequencies (0.1-100 kHz) in rather bulky ($10^{-2}$-$10^{-3}$ m) structures.[244–246] GHz applications have been demonstrated in one dimensional multilayer systems[182,247] while two dimensional systems have been mainly limited to theory.[243,248,249] Nevertheless, as the acoustic problems are scale-invariant, sooner or later, macroscopic phononic crystals will be replicated at the nanoscale. At the nanoscale, the principal existing challenge is the selective excitation and detection of acoustic waves that possess these special aforementioned properties. Towards this goal, Brendel et al.[243] suggested the use of the radiation pressure in an optomechanical cavity. In this configuration, it is relatively simple to change and control the optical power that couples into the optomechanical cavity. The radiation pressure can then be carefully controlled such that the desired acoustic waves can be effectively generated and launched. Later, the readout can be measured by detecting the sidebands of the reflected laser beam used to excite the acoustic wave. Another possibility is the use of IDTs as "acoustic wave launchers".[250] This platform consists of two IDTs coupled to an acoustic waveguide. One of the IDTs focuses the acoustic power at one end of the waveguide (emitter) while the second one is positioned at the other end of the waveguide (receiver). This particular arrangement could allow for the generation and detection of acoustic phonons travelling through a topologically protected waveguide.[251] Simply by reversing the emitter-receiver configuration, it is possible to demonstrate the unidirectionality of a topologically protected acoustic waveguide.

Several scientific advancements have shown the increasing relevance of phononics for information and communication technologies (ICTs) such as resonant acoustic nanocavities, surface phonon polariton lasers, micro-/nano-electromechanical devices, and non-diffusive thermal transport, all of which open the way for manipulating phonons as information carriers. The concept of topologically protected states has enormous potential for information and communication technology as these states can transport energy without dissipation. Similarly, the acoustic version of a topological insulator will allow for phonon transport along surface or edges but not within the bulk. A topological phonon is also immune to scattering by defects and other perturbations (i.e., they can travel along a surface without backscattering). Such technology could enable a platform for robust waveguides, improved acoustic-based devices (e.g., mobile phone sensors, touchscreen, gas, mass and pressure sensors among others), and radio (RF) and intermediate frequency (IF) filters.[252] The use of phonons as information carriers opens the possibility of low power computation processes, which would help to reduce energy consumption. Much more generally, robust control over phonon coherence, dynamics, and mean free path promises access to the tunability of all material properties that are dependent on the phonon physics that we have mentioned throughout this review.

**Funding and Acknowledgements:** We acknowledge the support from the project LEIT funded by the European Research Council, H2020 Grant Agreement No. 885689. ICN2 is supported by the Severo Ochoa program from the Spanish Research Agency (AEI, grant no. SEV-2017-0706) and by the CERCA Programme / Generalitat de Catalunya. R.C.N. acknowledges funding from the EU-H2020 research and innovation programme under the Marie Sklodowska Curie Individual Fellowship (Grant No. 897148). A.E.S. acknowledges support by the H2020-MSCA-IF project THERMIC-GA No. 101029727. F.C. acknowledges funding from the scholarship BES 2016 077203 granted by the Spanish government ministry of economy, industry, and competitiveness. P.X. and M.S. acknowledge funding from the H2020-FET project NANOPOLY (Grant No. 289061). P.X. additionally acknowledges support by a Ph.D. fellowship from the EU Marie Skłodowska-Curie COFUND PREBIST project (Grant No. 754558). OF is supported by BIST PhD fellowship Horizon 2020 research and innovation programme under the Marie Sklodowska-Curie grant agreement No. 754558.

**Author Contributions:** R.C.N and E.C.-A. designed, reviewed and edit the whole manuscript; R.C.N. wrote and edit IDT and OM-based sections; A.E.S. wrote and edit AFM-based section; F.C., O.F. and M.S. wrote and edit BLS-based section; M.G. wrote and edit LDV-based section; M.P. wrote and edit P&P-based section; J.J. MG wrote and edit IDT and P&P-based section; P.X. and E.C.-A. wrote and edit Raman-based section; Resources and funding adquisition C.M.S.T; R.C.N., A.E.S., C.M.S.T. and E.C.-A supervised the work. All authors have read and agreed to the published version of the manuscript.
**Conflicts of Interest:** The authors declare no conflict of interest.

**ORCID ID:**
Ryan C. Ng: 0000-0002-0527-9130
Alexandros El Sachat: 0000-0003-3798-9724
Francisco Cespedes: 0000-0001-5444-7113
Juliana Jaramillo-Fernandez: 0000-0002-4787-3904
Peng Xiao: 0000-0002-4711-2566
Omar Florez: 0000-0001-6662-9811
Marianna, Sledzinska: 0000-0001-8592-1121
Clivia Sotomayor-Torres: 0000-0001-9986-2716
Emigdio Chavez-Angel: 0000-0002-9783-0806